# Demonstration of synergic Fresnel and Fraunhofer diffraction for application to micrograting fabrication

Pritam P Shetty, Jayachandra Bingi*

Bio-inspired research and development (BiRD) laboratory, Photonic Devices and Sensors (PDS) Laboratory, Indian Institute of Information Technology Design and Manufacturing (IIITDM) Kancheepuram, Chennai, India - 600127.

Email: phy18d005@iiitdm.ac.in , bingi@iiitdm.ac.in

--------------------------------------------------------------------------------

## TABLE OF CONTENTS

--------------------------------------------------------------------------------

## Abstract

Diffraction is a manifestation of light at edge due to its wavelike nature. The well-known diffraction phenomena are Fresnel and Fraunhofer, they find variety of applications individually. But the synergy of two phenomena is not studied and understood, which is important to understand the compound optical instruments. This research studies and demonstrates the synergic patterns of Fresnel and Fraunhofer diffractions. The combined diffraction resulted in the patterns similar to Hermite-Gaussian Beam intensity distribution in both simulated and experimental results. Further, the combined phenomena is implemented in speckle lithography to fabricate wide area micro grating patterns which are better than simple speckle diffraction grating patterns. This work may contribute to the deeper understanding of complex diffraction in optical instruments and also shows the way for robust fabrication of micro gratings.

## Introduction

Diffraction is a phenomenon that occurs when light encounters an edge obstacle. There are two types of diffraction called Fresnel and Fraunhofer diffraction, observed at near-field and far-field region. The study of Fraunhofer diffraction has allowed its application in Plasma turbulence measurements[1], particle sizing[2], vibration sensing, metrology[3] etc. Similarly Fresnel diffraction studies paved the way for its applications in superfocusing for microscopy[4], characterization of spectral line[5], thin film thickness measurement[6] etc.

In most of the optical experiments and instruments, the circular aperture and vertical slits are used. The general assumption in these experiments is that the illuminating light is a plane wave. But, in the cases where beam divergence is involved the illuminating light can be spherical wave. In such cases the propagation of light may not only result in Fraunhofer diffraction but also the Fresnel. The Fresnel diffraction is more probable due to spherical wave illumination where we can observe the Fresnel pattern beyond the near field. This is generally considered as noise or mostly unnoticed. We present one such case here in Laser diode module where circular aperture results in Fresnel diffraction followed by Fraunhofer diffraction at vertical slit. To address this issue there is a need to study for combined effect of Fresnel and Fraunhofer diffraction. But the combined Fresnel and Fraunhofer diffraction pattern has not been studied.

On the other hand, the speckle lithography[7] is a method where the grating patterns are formed from individual speckle diffraction. Exploiting the combined effects in speckle lithography could give better quality gratings than speckle gratings mentioned in the literature[8]. This compound diffraction patterns can also be used for fabricating nanostructures like metasurfaces, anti-reflective surfaces[9][10][11] etc. Hence this paper focuses on study of combined Fresnel and Fraunhofer diffraction patterns and their implementation in speckle lithography for fabricating wide area micro grating patterns

## Results and Discussions

**Demonstration of Synergic diffraction:**

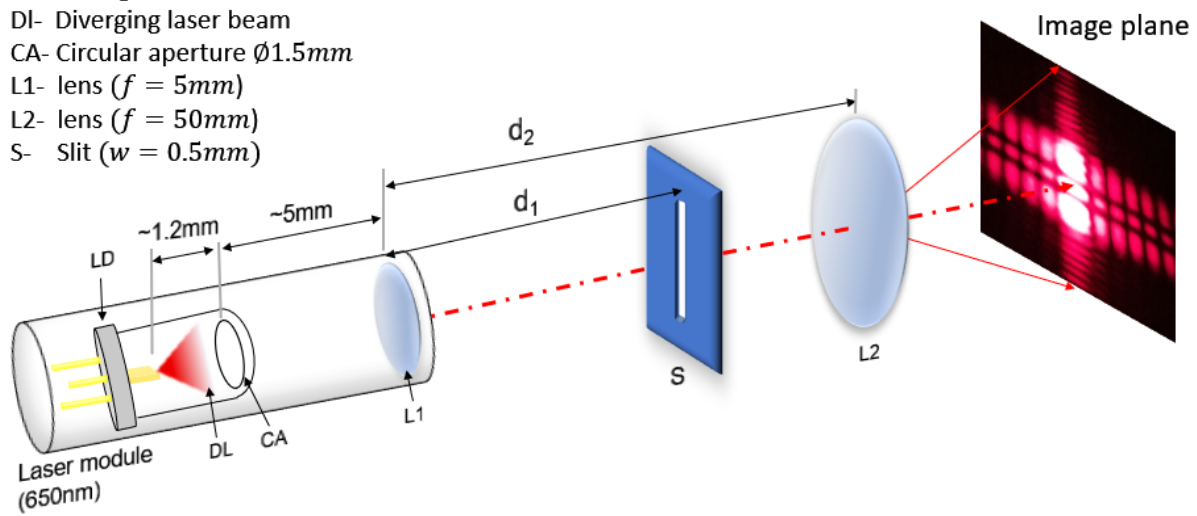

*Figure 1: Experimental setup for synergic diffraction phenomenon.*

Figure 1 shows the experimental setup involving slit, imaging lens (L2) and laser diode module (650nm). After investigating internal components of laser module, we noticed a circular aperture (CA) embedded in design of laser diode. Fresnel diffraction occurred due to illumination of CA by highly diverging beam from laser diode. Lens L1 in LD module collimates the Fresnel diffracted pattern from CA. As the laser source is highly diverging beam as shown in figure 1, lens L1 allows long distance propagation of this Fresnel diffracted field. Beam size before slit S is around ~5mm, which is Fresnel diffracted field before it enters slit S.

To demonstrate the Fresnel to Fraunhofer transition of diffracted field from CA, an experiment as shown in figure 2(a) is performed. Here a CA of 1mm diameter is illuminated by diverging laser beam. Fresnel to Fraunhofer transition of concentric ring intensity pattern is observed when imaging lens is moved away from CA. It is clear from the figure 2(a) that at less than 60 cm lens-CA distance the Fresnel diffraction is prominent whereas after 60 cm Fraunhofer diffraction is obvious for the given setup.

For the input beam with divergence, the equation for Fresnel number becomes $F = \frac{a^2}{\lambda L} + \frac{a^2}{\lambda R}$ [12].

Where, a = aperture radius

λ= wavelength

L= distance between aperture and observation plane

R=radius of curvature of beam incident on aperture

Here if F>>1, Fresnel diffraction occurs at observation plane and when F<<1, Fraunhofer diffraction occurs.

As per the parameters of the experiment (figure 1) such as

aperture radius, $a$ =0.75mm

wavelength, $\lambda$=650nm

distance between aperture and observation plane, $L$=7mm radius of curvature of beam incident on aperture, $R$=1.42mm (considering the aperture to light source distance 1.2mm) which gives the Fresnel number =733 >>1 that results Fresnel diffraction at longer distances too.

Further, to prove the role and importance of divergent beam in the observed phenomenon the following experiment has been conducted.

The same experiment as shown in figure 2(a) continued with slit of 0.5mm width in front of lens. This made Fresnel diffracted beam to undergo Fraunhofer diffraction there by forming a compound diffracted beam as shown in figure 2(b). Further when the divergent beam is replaced with collimated beam, intensity distribution as shown in figure 2(c) is observed. This demonstrates that the role of diverging beam to achieve Fresnel diffraction is crucial where the plane wave approximation no longer works and beam could be a spherical wave.

Further, synergic patterns are studied with respect to slit - laser module ($d_1$) distance, imaging lens - laser module($d_2$) distance shown in figure 1. The synergic diffraction pattern variations are shown in figure 3. The number of fringes in vertical direction increased when distance $d_2$ decreased, this can be attributed to variation in fringes for Fresnel diffraction component in compound beam. Distance between fringes in horizontal direction reduced when $d_1$ increased. Hence Fraunhofer component of compound beam is affected by distance $d_1$.

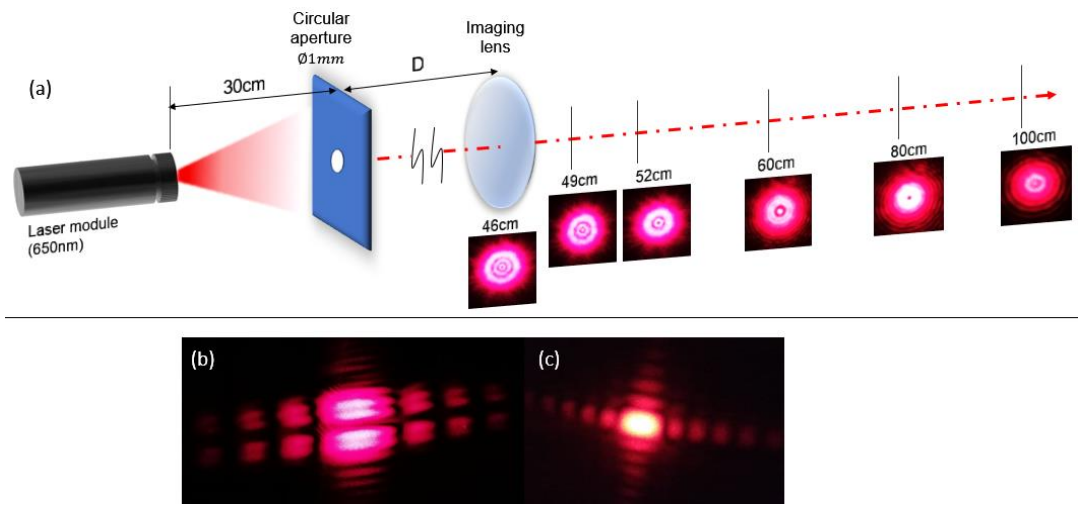

*Figure 2: a) Experimental Setup to demonstrate Fresnel to Fraunhofer Diffraction transition after CA as shown in inset images, b) shows the output image of beam when laser with spherical wave output is made to pass through CA and S arrangement. and c) shows output beam image for plane wave input to CA and S arrangement.*

Interestingly, the intensity distribution of the vertical set of fringes is strikingly similar to Hermite Gaussian (HG) modes (figure 4). But, consecutive lobes in ideal HG modes are 180° out phase from each other [13]. Moreover, HG beam can be converted into Laguerre Gaussian(LG) beams by using astigatic mode converter [14]. To prove the HG beam nature of the synergic diffraction pattern, it is allowed to pass through astigatic mode converter. It is observed that they are not converted to LG beam. This suggests that beams in figure 3 resembled HG beams only in terms of intensity distribution but not in terms of phase. Still the spatially varying amplitude of the beams generated in this research can be explored for optical traps as there is no complexity in beam generation as compared to HG beam generation. Figure 3 clearly shows that just by varying $d_1$ and $d_2$ the lobes can be tuned. Hence, successful creation of required phase difference among lobes may also give the scope for generation of mode tunable LG beam generators.

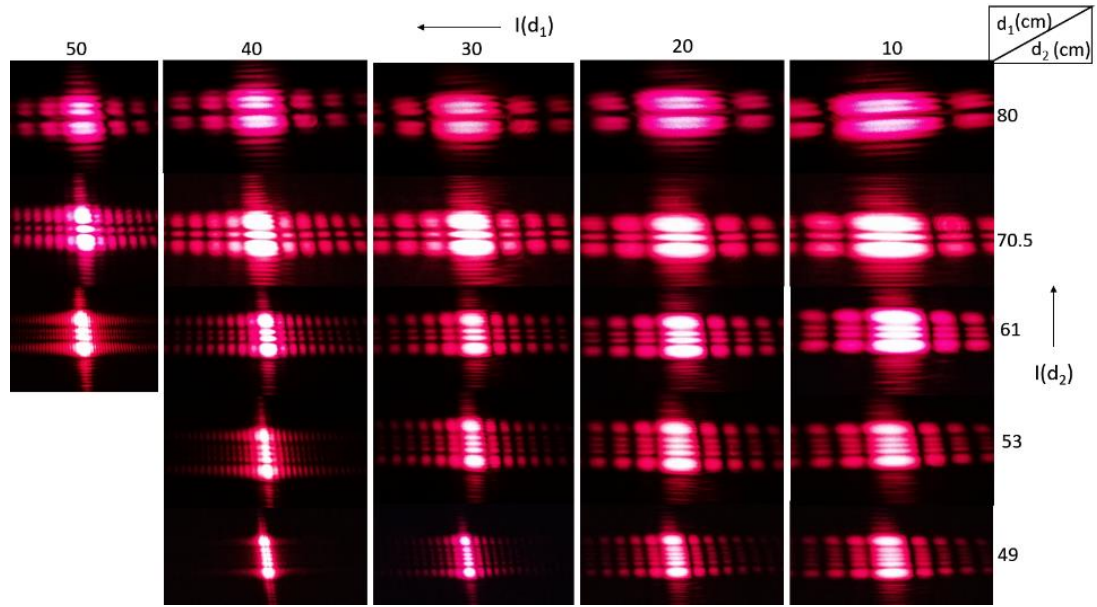

*Figure 3: Output Beams from the experimental setup as shown in figure 1 for variations in distances $d_1$ and $d_2$*

**Theoretical simulation of synergic diffraction:**

In the experiment (figure 1), the divergent beam is incident on CA and undergo Fresnel diffraction. The Fresnel pattern is collimated by lens L1 and incident on the slit which results in Fraunhofer diffraction, further magnified by another lens L2 before reaching imaging screen. The combined pattern due to successive Fresnel and Fraunhofer diffractions is observed on the screen. In the Literature, theoretical works are done for pure Fresnel and Fraunhofer diffractions separately with plane and spherical waves as incident field. But Fraunhofer with Fresnel pattern as incident field is not attempted theoretically. Which is complex and need separate dedicated work.

Keeping these constraints in mind the attempt is made to explain the experimental results shown in figure 3 by using the numerical simulations of general theory of diffraction i.e. by solving Fresnel-Kirchhoff integral. The software used to simulate is LightPipes library in python[15]. Due to limitations such as usability of only plane wave with small divergence and incorporation of lens, the simulations are done assuming plane wave undergoing two diffractions successively.

Figure 4 shows comparison for the beam generated from simulated and experimental results. Gaussian beam field of wavelength 650nm with small divergence is passed through CA of diameter 3mm and slit of width 0.5mm and then incident on screen. Number of fringes in vertical direction increased when distance between CA and imaging screen($d_2$) decreased and vice versa. The sharpness of the slit edges is not considered for simulation.

As the simulation is done to test the successive diffractions through aperture and slit, the outcoming compound diffraction patterns are compared with respect to number of nodes (singular dark regions). Considering the trend in node changes as a function of reduction in image plane-CA distance(d2) and pattern of intensity distribution especially at Node 3 and Node 4 it is clear that phenomena is due to successive diffraction of light. But the experimental values are 80cm 70.5cm, 61cm, 53cm whereas the simulation distances are 740cm, 505cm, 387cm and 311cm respectively for node 1 to node 4. This variation in distances is attributed to the non-usage of lenses in simulation. Also, each set of vertical fringes in compound diffracted field look very similar to standard Hermite-Gaussian modes as shown in figure 4.

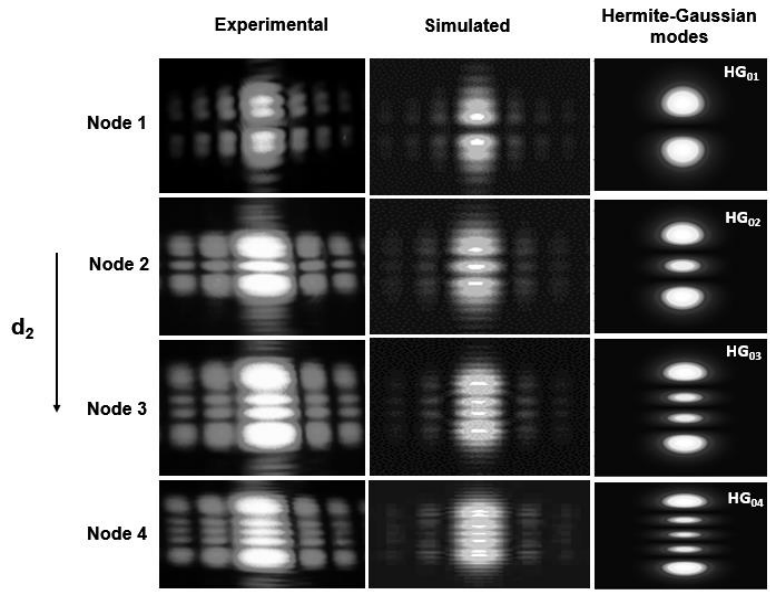

*Figure 4: First and second row shows as distance $d_2$ decreases number of nodes increases in both experimental and simulated results. The experimental values are 80cm 70.5cm, 61cm, 53cm whereas the simulation distances are 740cm, 505cm, 387cm and 311cm respectively for node 1 to node 4. Third row shows standard Hermite Gaussian modes.*

The observation of synergic diffraction patterns indicates that the grating vector of the pattern is in y-axis (Vertical) direction with the prominent intensity distribution across different orders. This is radically different from simple Fraunhofer diffraction where pattern is formed in z-axis direction usually with diminishing intensity at higher orders. This gives the advantage to fabricate the gratings with uniform fringe patterns. From literature [8] the speckle can undergo Fraunhofer diffraction and form a micro-grating pattern where intensity diminishes across the higher orders which reduce the quality of grating. Instead of the normal Fraunhofer diffraction we implement this compound diffraction phenomenon in speckle lithography to fabricate high quality micro gratings.

**Application to Speckle Lithography:**

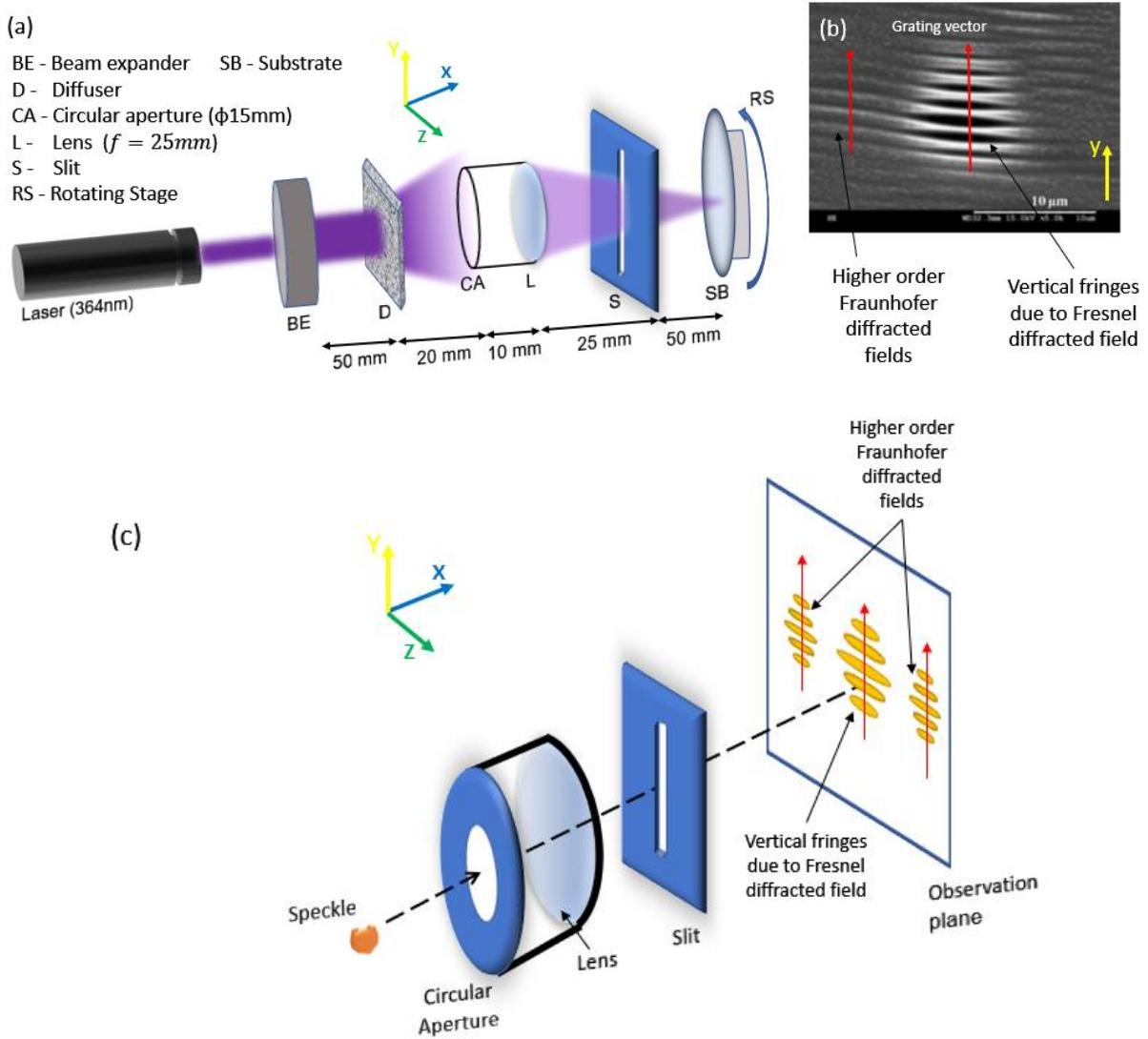

*Figure 5: (a) The speckle lithography experimental setup where the slit(S) width values are 1,1.5 and 2mm and the substrate is Si coated with AZ 7220 photoresist (b) the fringes in the individual speckle where the grating vector of fringes and slit length vector are parallel(Y-direction), (c) Interaction of speckle with apertures to undergo compound diffraction.*

Speckle pattern is a granular intensity pattern[16]. J. Bingi et. al. proposed and realized the phenomenon of individual speckle diffraction at the sharp edges which resulted in the formation of random grating structures[8]. The edge diffraction of an individual speckle is similar to Fraunhofer diffraction phenomena. The demonstrated synergic diffraction phenomena has strong vertical fringe pattern in zeroth and higher order Fraunhofer spots which is not usual simple diffraction. Further the vertical fringe pattern is similar to HG beam in the macroscopic experiment. To validate the phenomena as a general light behavior when it undergoes successive Fresnel and Fraunhofer diffractions, we adopted the speckle lithographic experiment where each speckle is tested to show the same phenomena at microscopic level.

In the experimental configuration, an ensemble of speckles is created using a 364 nm laser (coherent innova 90c). Further, the speckle beam (beam with many speckles) is allowed to diverge such that well-defined individual speckles are formed (field discontinuities become explicit). This speckle is allowed to pass through an optimally configured lens- aperture set up (figure 5(a)). Speckles that are coming through the slit are allowed to fall on the substrate coated with photoresist (AZ 7220/AZ9620) (thickness < 1µm coated on the substrate). The lens diameter and focal lengths are kept as 25 mm and 25 mm, respectively.

The slit width is varied from 2 mm to 1mm and the vertical length of the slit is kept as > 25 mm with open top and bottom ends. The slit is made of a metal sheet (aluminum) with a thickness of 0.5-0.6 mm. The substrate is placed at 1 cm from the aperture to record the speckles coming through the slit. The speckle in an ensemble is controlled by laser beam power, size and time duration of the incident laser. An optimized set of parameters such as 5 mW power, beam size of 2 cm and time duration of 500 msec are the experimental parameters considered apart from the optimized photoresist development time of 30-40 sec in this experiment, with MIF 826 developer.

The arbitrary shaped speckles generated from the ground glass are converged by the lens and pass through the aperture. When the aperture width is less than 2 mm, the formation of an intra-speckle fringe is observed. As demonstrated above, the intra speckle fringes are formed in the Y-direction, hence the fringes are named as V-fringes (vertical fringes) here. Figure 5(b) shows the SEM image of a speckle with distinguishable intra speckle fringes. Figure 5(c) describes the synergic diffraction by using speckles. Here the vertical fringes i.e. the grating vector is parallel to the length vector of slit which is the evidence for synergic phenomena. It is absolutely repeatable in different conditions.

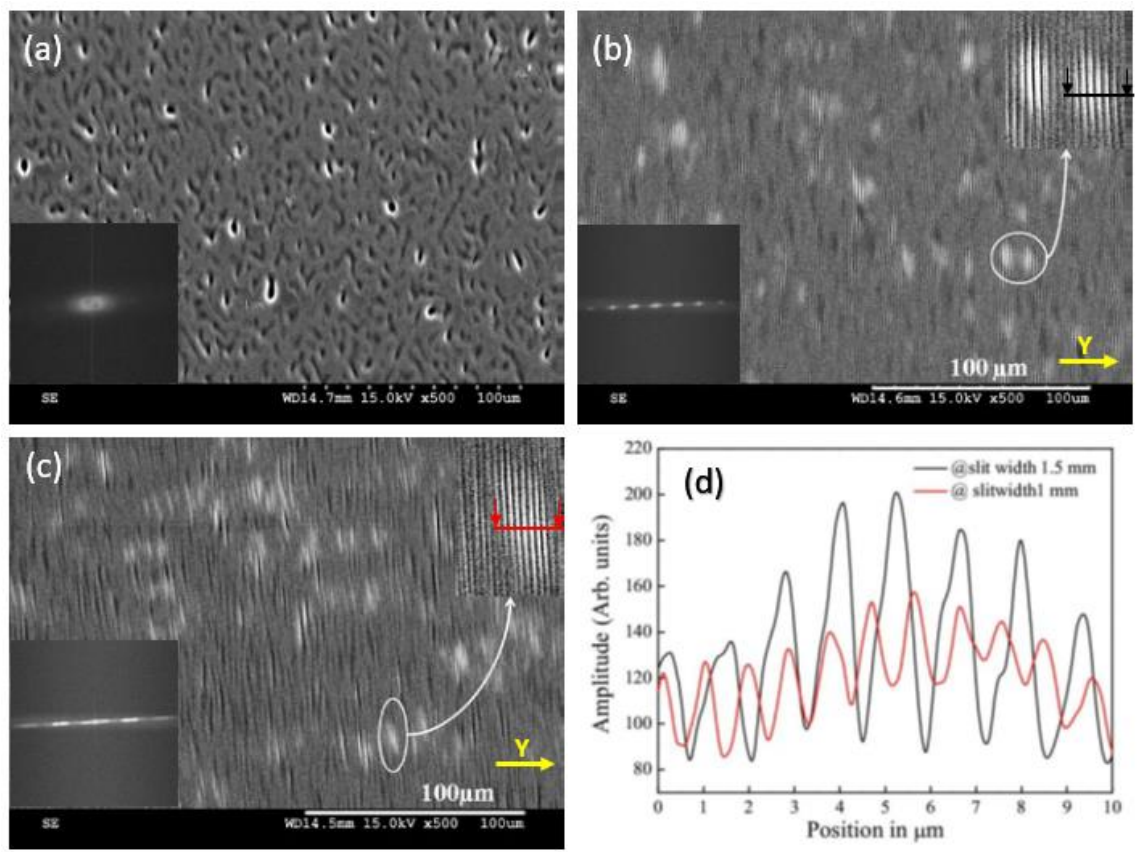

*Figure 6: (a, b, c) Speckles printed on the photoresist with varying slit widths 2 mm, 1.5 mm and 1mm respectively(Bottom right inset-FFT spectrum), (d) the plot showing the fringe period variation with slit width(Top right inset in (b) and (c) shows section lines).*

Further, the experiment is carried out with different slit sizes of 2 mm, 1.5 mm and 1 mm. Figure 6(a), (b) and (c) shows the respective SEM images of the speckle micro grating recorded on photoresist for different slit widths. The V-fringes are not formed at 2 mm slit width whereas the well-defined V-fringes are observed with 1.5 and 1 mm slits respectively as shown in figure 6(b), (c). The top right inset shows the detailed view of individual speckle with V-fringes. It is clear from figure 6(a), (b) and (c) that the individual speckle is getting elongated when slit width is reduced from 2 mm to 1 mm, in Z-direction. At the same time the V-fringe period is also changing with the elongation. The bottom right inset shows Fourier transform frequency spectra for the SEM images. As the speckle patterns are random in distribution the generated grating patterns also distributed randomly across the area. Fourier transform frequency spectra indicates that at 2mm slit width there are no gratings patterns formed on the photoresist which is indicated by a ring like pattern at low frequency region in the FT spectrum. Further at slit widths 1.5mm and 1mm different frequency components observed in the FT spectrum which indicates the distribution and variation in periodicities is in one direction.

Figure 6(d) shows the plot of V-fringe periods recorded at 1.5 mm and 1 mm. It is clear from the plots that period of V-fringe pattern within the individual speckle is reducing at 1 mm slit width. Using 354nm wavelength it is possible to achieve submicron level feature size.

Maximum fabrication area is 10x10mm. Usage of lower wavelengths such as 256nm laser speckles could further reduce the resolution. Further, at a time fabrication area is equal to area of aperture and imprinting time on the photoresist is in the order of seconds hence large are fabrication is also possible

## **Conclusion**

In summary this work showed the combined effect of Fresnel and Fraunhofer diffractions and indicated that the intensity distributions are significantly different from simple Fresnel diffraction both by simulation and experimental results. The compound phenomena is implemented in speckle lithography to fabricate the micro gratings with sub-micron pitch. It is also showed that the gratings pitch and area can be modulated by varying the experimental parameters such as slit width, aperture diameter etc. this work demonstrated the significance of compound diffraction effects may be useful in understanding the optical instruments better. At the same time implementation of this compound phenomena in speckle lithography opens the opportunity for robust fabrication of wide area micro gratings.

Acknowledgements: We acknowledge the funding support from DST India under INT/RUS/RFBR/P-262. Authors also acknowledge the facility support from Prof. Murukeshan V M from NTU, Singapore.

COI: Authors declare the no conflicts of interest.